# PHASE PLANE ANALYSIS OF THE PHOTOMETRICAL VARIATIONS OF LONG-PERIOD VARIABLES

Kudashkina L.S., Andronov I.L.

Department of Mathematics, Physics and Astronomy,
Odessa National Maritime University, Odessa, 65029 Ukraine
*kuda2003@ukr.net, tt_ari@ukr.net*

**ABSTRACT.** Using the phase plane diagrams, the phase light curves of a group of the Mira-type stars and semi-regular variables are analyzed. As generalized coordinates $x$ and $\dot{x}$, we have used $m$ – the brightness of the star and its phase derivative. We have used mean phase light curves using observations of various authors. The data typically span a large time interval (nearly a century). They were compiled from the databases of AAVSO, AFOEV, VSOLJ, ASAS and approximated using a trigonometric polynomial of statistically optimal degree. As the resulting approximation characterizes the auto-oscillation process, which leads to a photometrical variability, the phase diagram corresponds to a limit cycle.

For all stars studied, the limit cycles were computed. For a simple sine-like light curve, in e.g. , $L_2$ Pup, the limit cycle is a simple ellipse. In a case of more complicated light curve, in which harmonics are statistically significant, the limit cycle has deviations from the ellipse.

In an addition to a classical analysis, we use the error estimates of the smoothing function and its derivative to constrain an "error corridor" in the phase plane.

**Keywords:** Mira-type stars, semi-regular variables, limit cycle, phase plane diagrams.

## 1. Introduction

The pulsations of long-period variable stars – Mira-type, semi-regular variables studied are significantly worse than, for example, of the cepheids, because of the difficulties associated with the complex interaction of pulsations and convection, and strong nonlinear effects, leading to the formation of shock waves and the mass loss, with the problems of the transfer of the radiation in a cool extended atmospheres with a high degree of non-adiabatic pulsations due to the commensurate dynamic and thermal time scales for these stars.

The dynamic time scale $\tau_h$ characterizes the rate of change of parameters of stars with the motions with velocities comparable to the velocitie of sound $v_s$. The order of magnitude $\tau_h \propto \dfrac{R}{v_s}$, where $R$ – the characteristic size of the star. For equilibrium star this time is order to time of free fall:

$$t \propto R^{3/2}/\sqrt{GM}. \qquad (1)$$

From these considerations, for example, in the work (Kudashkina & Marsakova, 2013) for stars R Aql, R Hya and T UMi, the time of the change of pulse period has been estimated "to zero" at the time of compression of the layers of the star due to the termination of helium burning in the layered source. Also the radius was estimated, from which begins the compression, for a range of stellar masses 0.9-1.5 $M_\odot$.

The thermal time scale $\tau_{th}$ determines the rate of cooling or heating of the star. When cooled in the absence of nuclear burning

$$\tau_{th} \propto \frac{U}{L} \propto \frac{GM^2}{RL}, \qquad (2)$$

since the energy $U$ the order of the gravitational energy of a star.

On the asymptotic giant branch (AGB), to which belong the investigated star, the star consists of a degenerate carbon-oxygen core and two layer sources (helium and hydrogen), positioned very close to each other. Above them is an extended hydrogen envelope.

The small thickness of the layered sources cause thermal flashes. Here, the stars are divided into two stages: early (EAGB) is the time interval between the end of helium burning in the core and the first thermal pulse of the helium layer source, and "thermally pulsing regime of the He-burning shell" (TPAGB) is the thermally pulsed burning mode in the helium layer. At the stage of a TPAGB the star is getting brighter in $M_{bol}$. Some theoretical calculations show that, for example, zirconium stars over a range of luminosity and temperature correspond to the early stages of AGB when helium burning occurs stationary. At the same time, several observational features say, rather, in favor of finding these stars in more evolved evolutionary stage on the AGB (Kudashkina, 2003).

The different studies show that long-period variables pulsate in a fundamental fashion and in the first overtone. The latter is especially for semi-regular variables, but a significant part of the Mira-type stars are the overtone pulsators. In particular, in the article of Fadeev (1993) is that EAGB stars represent a homogeneous group of fundamental pulsators with periods from 10 to 400 days, while the TPAGB stars show different pulsation proper-ties, depending on the initial mass. So for a range of masses from 2 to 3 solar the ultimate period of the first overtone of about 300 days, and 430 days, respectively for 2 and 3 $M_\odot$. Depending on the initial mass and surface luminosity, these stars pulsate in the fundamental mode, either in the first overtone. Some objects show variability of either optical, or maser emission (e.g. Kudashkina & Rudnitskij (1988).

Indirect dependence of the pulsation properties from surface luminosity is also seen in the various photometric dependencies. For example, the three-dimensional diagram of the photometric parameters: the period ($P$), the amplitude of light ($\Delta m$), the slope of the ascending branch of the light curve ($dm_i/dt$) were obtained (Kudashkina (2015).

The most significant correlation shows the dependence of the slope of the ascending branch of the period and amplitude.

$$dm_i/dt = -\underset{\pm 0.0075}{0.0633} + \underset{\pm 0.000021}{0.000165}\,P - \underset{\pm 0.00134}{0.00642}\,\Delta m. \quad (3)$$

This study is the next step in the study of the photometric parameters of long-period variables – Mira-type stars and the semi-regular and relative objects – which can be used as an additional criterion the classification of these stars to the EAGB and TPAGB stages.

**2. Results**

It is known that the mode of oscillation corresponds to a periodic limit cycle, i.e. a closed phase trajectory to strive for all of the close trajectory over time. We calculated limit cycles for 19 long-period variables and 5 stars of RV Tau-type, based on the mean light curves, averaged over a long period of time. As generalized coordinates of the phase plane are taken $m$ – brightness of the star and its phase derivative. That is, the curve of evolution of the brightness of the star in the phase space $m = \Phi(\varphi, m_0)$ is a solution of the equation

$$\frac{dm}{d\varphi} = F(m), \quad (4)$$

where $m_0$ is the value at $\varphi=\varphi_0$.

For approximation of the phase curve was used trigonometric polynomials

$$m(\varphi) = C_1 + \sum_{j=1}^{M}(C_{2j}\cos(j\varphi) + C_{2j+1}\sin(j\varphi)), \quad (5)$$

moreover, the optimal value of the degree $M$ is determined by the Fisher test with a critical probability of "false alarm" (FAP=False Alarm Probability) $10^{-3}$.

The phase derivative in radians calculated simple

$$\dot{m}(\varphi) = \sum_{j=1}^{M} j \cdot (-C_{2j}\sin(j\varphi) + C_{2j+1}\cos(j\varphi)), \quad (6)$$

That is, upon differentiation, the amplitude of the harmonic is multiplied by $j$. Usually this leads to a greater increase in the relative statistical error of the derivative compared to the signal.

The examples of the phase portraits of the stars made using an improved version of the program FDCN (Andronov, 1994) and are shown in Fig. 1-17. Also the mean light curves are shown for several stars. For the analysis, we have used visual observations from the AAVSO, AFOEV and VSOLJ international databases, the ASAS survey, as well as the original observations by Whitelock et al. (2000) and Maffei &Tosti (1999).

Theoretically, for sinusoidal oscillations, the phase portrait is an ellipse with different units of measurement on axes – magnitude and magnitude "the radian" or "the period". Deviations from the ellipse expected for non-sinusoidal waves, i.e. if the degree of the trigonometric polynomial $M >1$. The calculation is performed in the cycle phase, the results for $m$, $\dot{m}$ are displayed with "3σ" corridors of errors on both coordinates. Naturally, the extreme points on each coordinate line "corridors of errors" for the other coordinate converge to a point like the letter Ж. Therefore, as the corridor of errors, use of external (and internal) part of the corridors of the errors for the two variables.

The list of stars and the basic data are given in table 1.

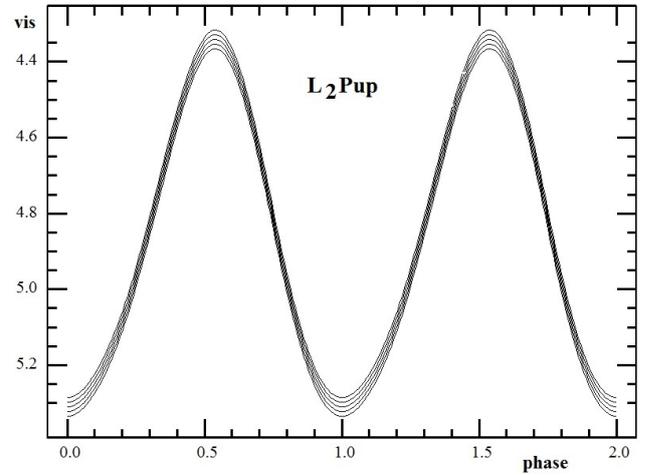

Figure 1: The mean light curve of $L_2$ Pup.

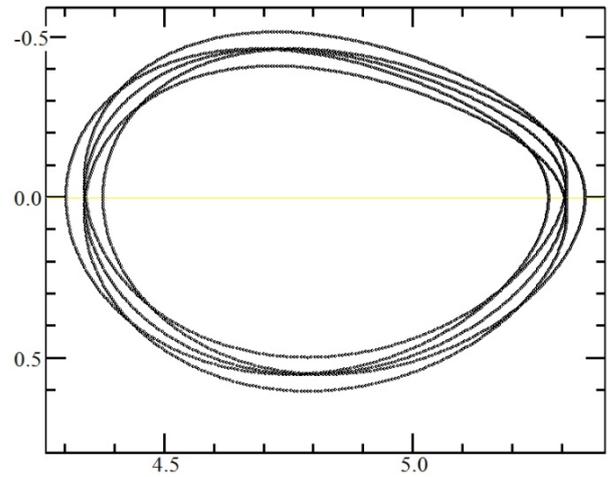

Figure 2: Phase portrait of $L_2$ Pup.

**4. Conclusion**

For the semi-regular variable S Per, as well as for $L_2$ Pup limit cycle is almost a perfect ellipse, despite the fact that the individual cycles of the oscillations are not quite regular. For Mira-type stars, with more regular oscillations, but a complex form of the light curve, limit cycles deviate strongly from the ellipse.

For U Mon (RV-type star), the phase portraits are shown for short and long cycles of variability. We can see, that the short-period cycle is more stable and regular then the long-period.

Table 1: The list of stars and their basic data

| N | Star | Type | Spectral type | Period, d |
|---|------|------|---------------|-----------|
| 1 | S Car | M | M2-M3e | 150.07 |
| 2 | U Her | M | M6.5e-M9.5e | 406.74 |
| 3 | X Oph | M | K1IIIv comp | 332.37 |
| 4 | R Aql | M | M5e-M9e | 280.84 |
| 5 | S Ori | M | M6.5e-M9.5e | 414.46 |
| 6 | Omi Cet | M | M5e-M9e | 333.6 |
| 7 | S Scl | M | M7-M8IIIe | 367 |
| 8 | RR Aql | M | M6e-M9 | 390.78 |
| 9 | X CrB | M | M5e-M7e | 241.1 |
| 10 | U UMi | M | M6e-M8e | 325.9 |
| 11 | U Cyg | M | C7.2e-C9.2 | 465.49 |
| 12 | V Cyg | M | C5.3e-C7.4e | 421.27 |
| 13 | BG Cyg | M | M7e-M8e | 288.1 |
| 14 | AM Cyg | M | M6e | 371.9 |
| 15 | R Leo | M | M6e-M9.5e | 312 |
| 16 | $L_2$ Pup | SRb | M5e | 137.17 |
| 17 | S Per | SRc | M3Iae-M7 | 816.8 |
| 18 | W Hya | SRa | M7.5e-M9ep | 381.7 |
| 19 | Y CVn | SRb | C5,4J(N3) | 267 |
| 20 | RV Tau | RV | G2eIa-M2Ia | 78.73 |
| 21 | U Mon | RV | F8eIb-K0pIb (M2) | 92 |
| 22 | EP Lyr | RV | A4Ib-G5p | 82.95 |
| 23 | R Sge | RV | G0Ib-G8Ib | 71.57 |
| 24 | DF Cyg | RV | G5-K4I-II | 776.4 |

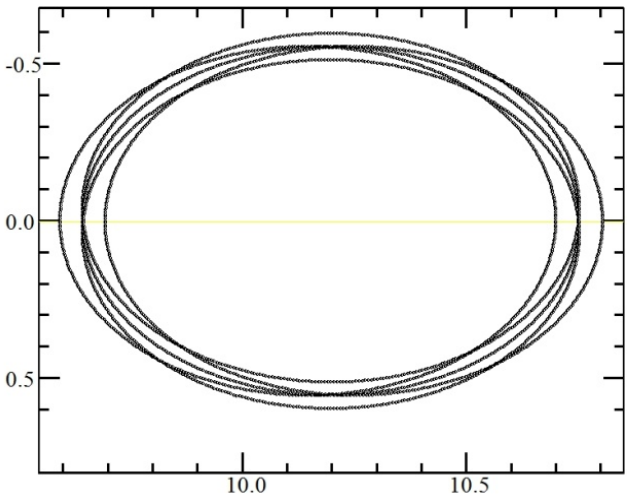

Figure 3: Phase portrait of S Per.

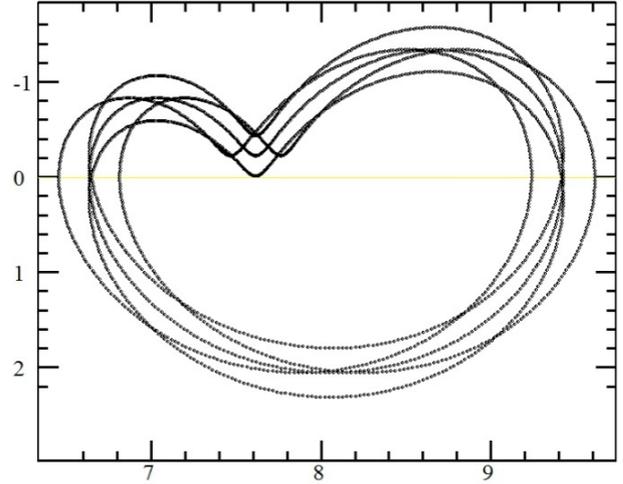

Figure 4: Phase portrait of W Hya.

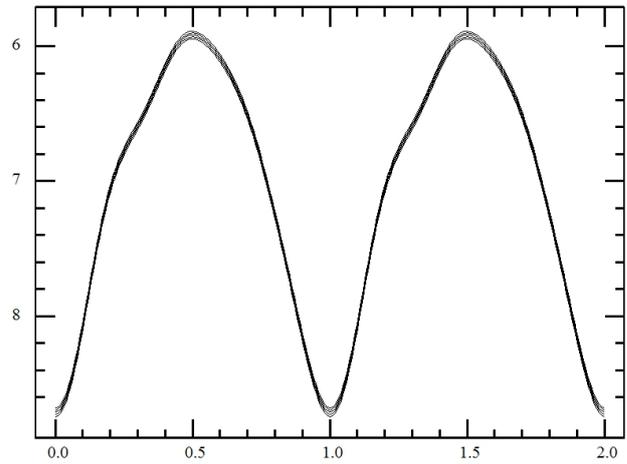

Figure 5: The mean light curve of S Car.

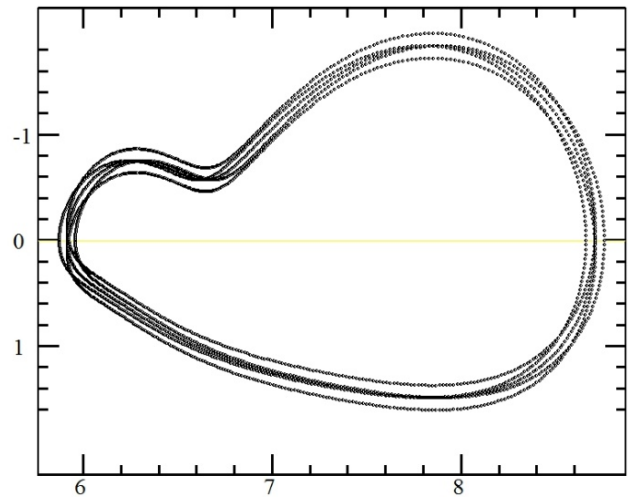

Figure 6: Phase portrait of S Car.

It was previously determined the value of $s$ – the degree of the trigonometric polynomial. It is equal to 1 for $L_2$ Pup (Kudashkina, 2016); 2 for S Car, R Aql (Kudashkina & Andronov, 1996), S Ori (Kudashkina, 2016); 5 for U Her (Kudashkina & Andronov, 1996).

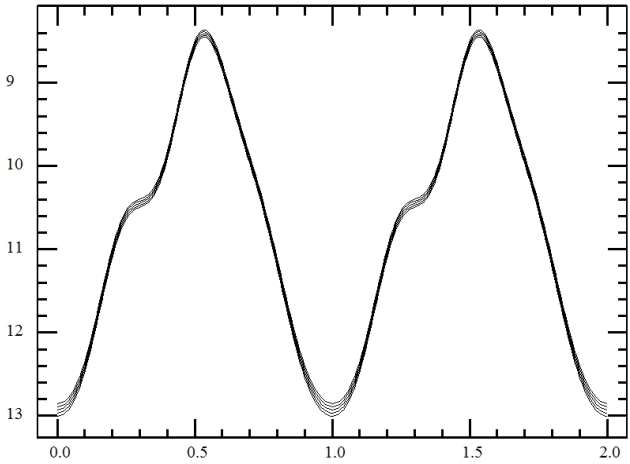

Figure 7: The mean light curve of S Ori.

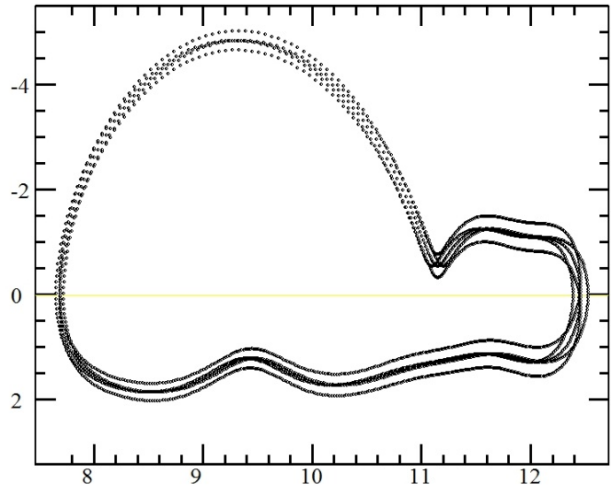

Figure 10: Phase portrait of U Her.

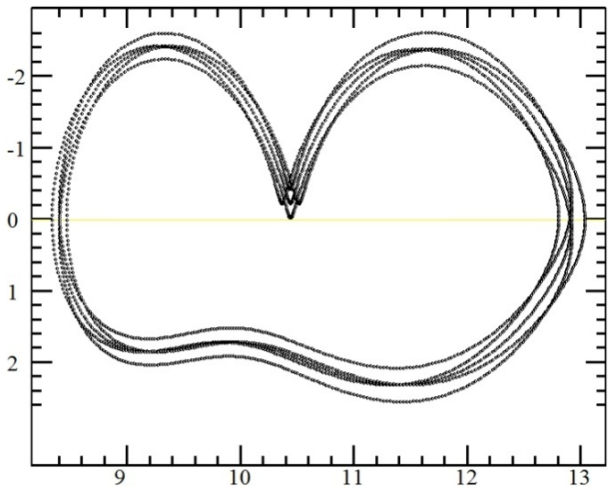

Figure 8: Phase portrait of S Ori.

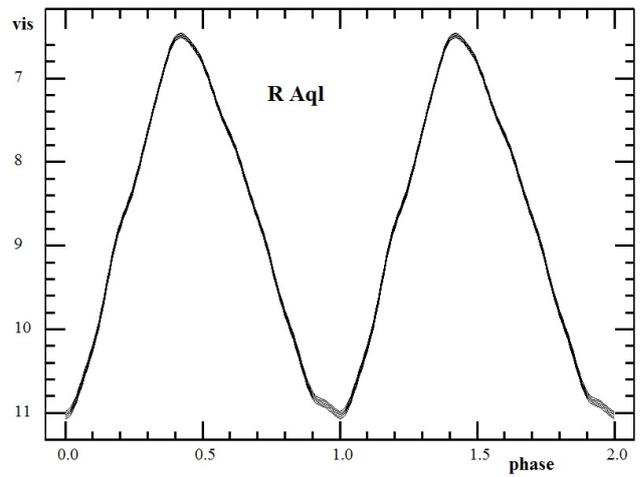

Figure 11: The mean light curve of R Aql.

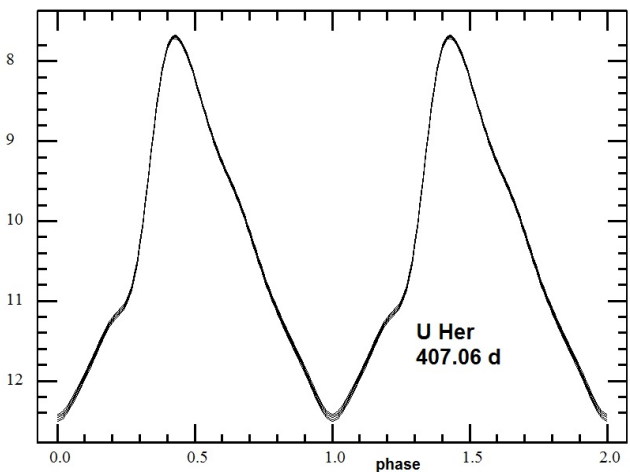

Figure 9: The mean light curve of U Her.

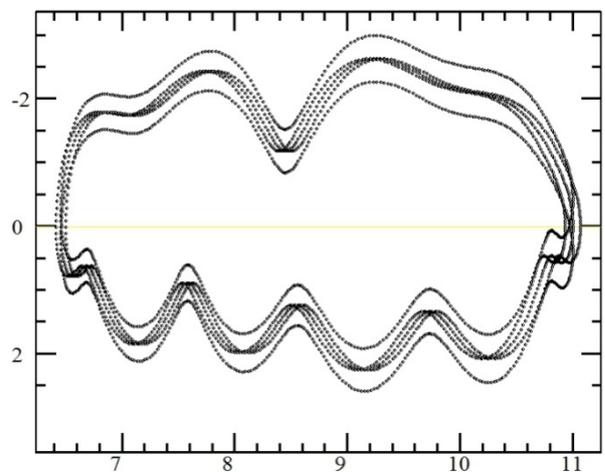

Figure 12: Phase portrait of R Aql.

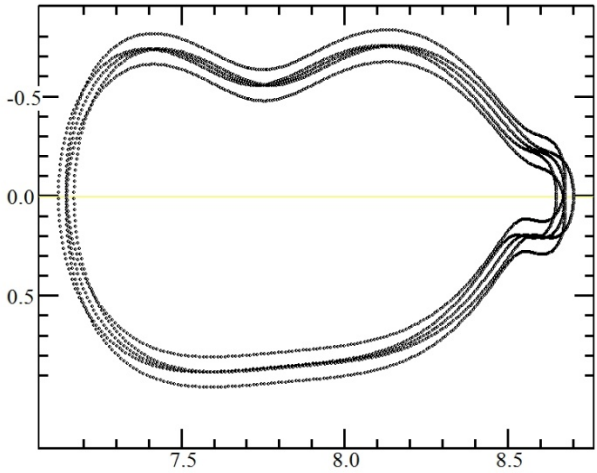

Figure 13: Phase portrait of X Oph.

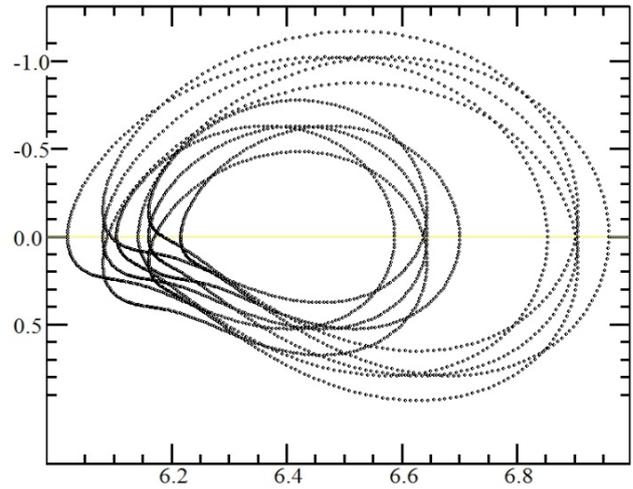

Figure 16: Phase portrait of U Mon.

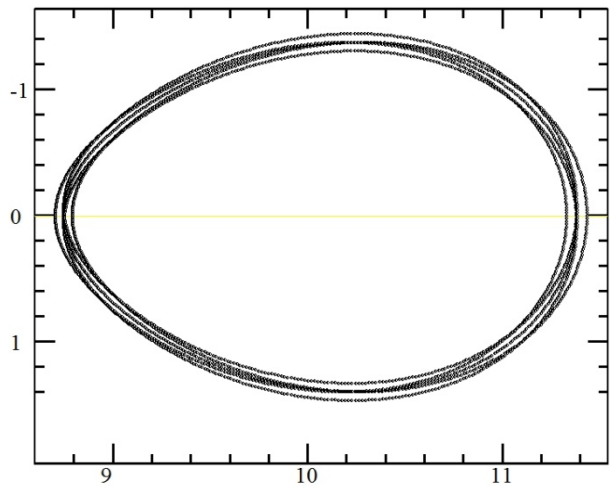

Figure 14: Phase portrait of U UMi.

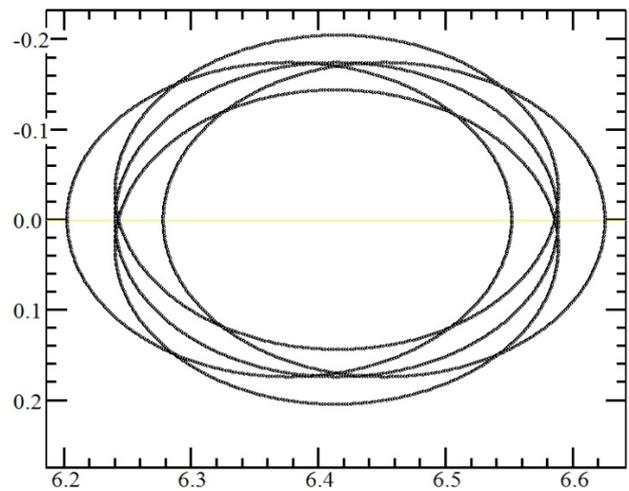

Figure 17: Phase portrait of U Mon with a half period.

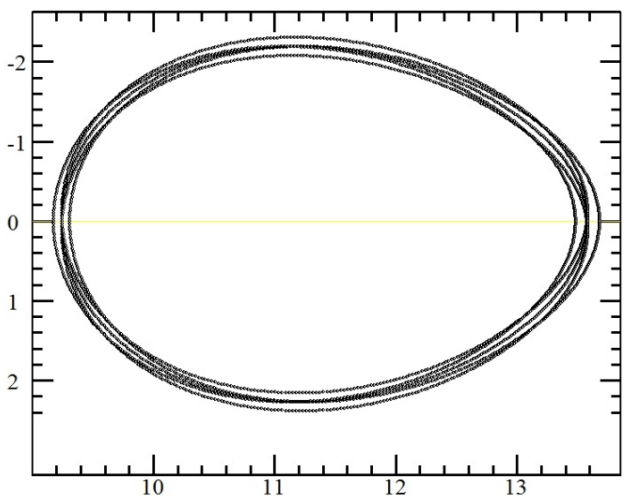

Figure 15: Phase portrait of X CrB.

*Acknowledgements.* This work is performed in the framework of the "Stellar Bell" part (pulsating variable stars) of the "Inter-Longitude Astronomy" project (Andronov et al., 2003, 2014, 2017). We acknowledge observations from the AAVSO (http://aavso.org), AFOEV (http://cdsarc.u-strasbg.fr/afoev), VSOLJ (http://www.kusastro.kyoto-u.ac.jp/vsnet/VSOLJ) and ASAS (http://www.astrouw.edu.pl/asas) databases.